\documentclass[pre,twocolumn,superscriptaddress,showpacs]{revtex4}
\usepackage{graphicx}
\usepackage{bm}
\begin{document}
\def\btt#1{{\tt$\backslash$#1}}
\def\tr{{\rm\,tr\,}}
\newcommand{\bra}[1]{\langle #1|}
\newcommand{\ket}[1]{|#1\rangle}
\newcommand{\braket}[2]{\langle #1|#2\rangle}
\newcommand{\half}{\textstyle{\frac{1}{2}}}
\newcommand{\beq}{\begin{equation}}
\newcommand{\eeq}{\end{equation}}
\newcommand{\bdis}{\begin{displaymath}}
\newcommand{\edis}{\end{displaymath}}
\newcommand{\bea}{\begin{eqnarray}}
\newcommand{\eea}{\end{eqnarray}}
\newcommand{\barr}{\begin{array}}
\newcommand{\earr}{\end{array}}
\newcommand{\beas}{\begin{eqnarray*}}
\newcommand{\eeas}{\end{eqnarray*}}
\title{Effect of conformations on charge transport in a thin elastic tube}
\author{Radha Balakrishnan}
\affiliation{The Institute of Mathematical Sciences, Chennai  600 113,
India}
\email {radha@imsc.res.in}
\author{Rossen Dandoloff}
\affiliation{Laboratoire de Physique Th\'eorique et Mod\'elisation,
Universit\'e de Cergy-Pontoise, F-95302 Cergy-Pontoise, France}
\email{rossen.dandoloff@ptm.u-cergy.fr}


\begin{abstract}
We study the effect of conformations
on charge transport in  a thin elastic tube.
Using the Kirchhoff model for a tube with any given Poisson ratio,
cross-sectional shape and intrinsic twist,
we obtain a class of exact solutions for its
conformation.
The tube's  torsion is found in terms of  its intrinsic twist and its
Poisson ratio, while its  curvature satisfies a nonlinear differential
equation which supports exact {\it periodic} solutions in the form of
Jacobi elliptic functions,
which we call  {\it conformon lattice} solutions.
These solutions
typically  describe  conformations with loops.
Each solution induces  a corresponding  quantum effective {\it periodic}
potential in the Schr\"{o}dinger equation for an  electron in the tube.
The wave function describes the delocalization of the electron
along the central axis of the tube.
We discuss some possible applications of this  novel mechanism
of charge transport.

\end{abstract}
\maketitle

\maketitle
\section{Introduction}

Nanotubes  and nanowires have attracted considerable attention due to their
their potential technological applications \cite{baug}. Carbon nanotubes and DNA are well known
examples. One important problem is to
understand  their mechanical properties. Another is to  gain some insight into
the effect of quantum confinement on
electronic transport in such nanostructures.\\

One striking observation concerning the DNA molecule is that its central
axis can take on
various curved conformations. Similarly, nanotubes need
not be always straight.
Helix-shaped nanotubes have been
observed
by Volodin {\it et al} \cite{volo}.
One way to   theoretically model these twisted  thin tubes (wires)
would be to assume that they are elastic filaments  described
within the framework of the Kirchhoff rod model \cite{kirc}, derive the
various possible curved conformations that can arise from it,
and see if they are in accordance with the shapes actually
observed. 
This would also help in studying the elastic properties
of these structures within this model. \cite{fons}
The natural question that arises
is how the  confinement of a particle to a curved path would affect
its transport.\\

Now, the problem of quantum transport of a free
particle in a thin curved tube has been studied by several authors
\cite{daco, gold, clar}.
They have shown that the curved geometry of the tube
essentially induces a potential in the path of the particle,
thereby affecting its  transport properties.\\

In a recent paper \cite{jpha}, we had analyzed the case when the thin
tube is made of elastic material, and applied it to electron transport
in a  biopolymer.
 We analyzed the
statics and dynamics of the thin elastic tube
using the Kirchhoff
 model\cite{kirc}.
Under certain conditions, the curvature function of the tube supports
a {\it spatially localized} traveling wave solution called
a  Kovalevskaya  solitary wave \cite{cole, gor1}.
We showed that this solution  corresponds to a
conformation which is a  localized twisted loop
traveling along the tube \cite{jpha}.
The localized bend
 induces a  quantum  potential well in the Schr\"{o}dinger
 equation for an electron in the tube \cite{daco,
gold, clar}, which  traps it
in the twisted loop , resulting in its efficient motion,
without change of form,
along the polymer.  Our result formalizes the
concept of a {\it conformon}  that has been
hypothesized in biology. \cite{volk, scot, keme}\\

The motivation for the present paper is as follows.  Firstly,
 the analysis of the Kirchhoff model given in \cite{jpha,gor1}
had assumed that the elastic tube had no intrinsic twist. It was therefore
not strictly applicable to either to DNA (which has  an intrinsic twist
of  about $10 \deg/$\AA, even in the straight relaxed state) or to
the class of intrinsically twisted  nanotubes.
In what follows, we present
a more general analysis of the Kirchhoff model,   by incorporating
an intrinsic twist in the tube, and
 point out the nontrivial role that it plays in
 determining the geometrical shape of the  axis of the tube.
Further, our analysis is valid for any  general value of the
Poisson ratio.
 Secondly, we show that the nonlinear differential equation for the curvature
function  obtained from the static Kirchhoff equations can, in addition to the
localized solution
discussed in \cite{jpha},  also support a class of {\it spatially periodic}
solutions in the form of Jacobi
elliptic functions.  We call them {\it conformon lattice solutions}.
 Each such solution induces a  quantum periodic potential.
Finally, we show explicitly that the Schr\"{o}dinger equation supports
an exact solution, which corresponds to
the electron getting  {\it delocalized}  along the axis of the
quantum wire,
 even in the static case.
 Thus this mechanism  for  electron transport
in a quantum  nanotube is distinct from  the conformon mechanism
discussed in  Ref. \cite{jpha}.

\section{The Kirchhoff model}

  We consider a thin elastic rod \cite{kirc}
whose central axis is  described by a
space curve $\mathbf{R}(s,t)$.
Here, $s$ denotes the arc length of the
curve and $t$ the time.
Let the  rod remain  inextensible with time evolution, with
the unit tangent to its axial curve  given by
${\bf t} =  {\bf R}_s$.
 (The subscript $s$ denotes the derivative with respect to
$s$). In the plane perpendicular to ${\bf t}$,
we define as usual \cite{struik}, two orthogonal unit vectors
 ${\bf n}$ and ${\bf b}$,
where the principal normal  ${\bf n}$  is  the unit vector
along  ${\bf t}_{s}$, and the binormal ${\bf b} = ({\bf t} \times {\bf n})$.
The rotation of the Frenet frame $({\bf t}, {\bf n}, {\bf b})$ as
it moves along the curve,
is given by the well
known Frenet-Serret equations, ${\bf
t}_{s} = k{\bf
n}$, ${\bf n}_{s} = -k {\bf t} + \tau {\bf b}$
and ${\bf b}_{s} = -\tau {\bf n}$, where
the curvature $k=|\mathbf{t}_{s}|$ and
the torsion
$\tau=\mathbf{t}~\cdot~
(\mathbf{t}_{s}~\times~\mathbf{t}_{ss})~/~k^{2}$.

For elastic tubes, instead of ${\bf n}$ and ${\bf b}$,
it is  more natural to use  `material'  orthogonal unit
vectors  ${\bf d}_1$ and ${\bf d}_2$ lying on the
cross-section of the tube,
perpendicular to its axial curve. For e.g., they could lie along the
principal axes of inertia  of the cross-section. Let
${\bf d}_{1}$  make an angle
$\phi$ with ${\bf n}$.
Hence we get  ${\bf
d}_{1} = {\bf n }
\cos \phi + {\bf b} \sin \phi $ and ${\bf d}_{2} = -{\bf n} \sin
\phi + {\bf b} \cos \phi$. Denoting the tangent ${\bf t}$
by ${\bf d}_{3}$, and using the Frenet-Serret
equations, it is easy to show that the
rotation  of the `material' frame  $({\bf d}_1, {\bf
d}_2, {\bf d}_3)$ along the curve is given by
\bea
 \mathbf{d}_{i,s}=\mathbf{k} \times \mathbf{d}_i,
\label{FE}
\eea
where $i=1,2,3$, and  the  vector $\mathbf{k}$
 is given by
\bea
\mathbf{k}= k_1\mathbf{d}_1 + k_2\mathbf{d}_2 +k_3 \mathbf{d}_3.
\label{k}
\eea
Here
\bea
(k_1, k_2, k_3) = (k\sin\phi,~ k\cos\phi,~ \tau + \phi_s).
\label{kcomp}
\eea

 Let the internal elastic force (or tension)
 and the total torque, that  act on each cross section of the rod be
given by $\mathbf{g}$ and $\mathbf{m}$ respectively. The Kirchhoff equations that
result from the conservation of linear and angular momentum at every point $s$
are given in dimensionless form by \cite{cole, gor1}
\bea \mathbf{g}_{s}= \mathbf{R}_{tt} \label{K1}\eea
and
\bea
\mathbf{m}_{s}+ \mathbf{d}_3 \times
\mathbf{g} = a \mathbf{d}_1 \times \mathbf{d}_{1,tt} + \mathbf{d}_2 \times
 \mathbf{d}_{2,tt}\,,
\label{K2}
\eea
where  the subcript $t$ stands for the derivative with respect to time.
The internal force is written in a  general form as
 \bea
\mathbf{g} =
g_1\mathbf{d}_1 + g_2 \mathbf{d}_2 + g_3\mathbf{d}_3\,.
\label{g}
\eea

We take into account the intrinsic twist $k_3^{0}$ of the elastic tube,
 by writing the  constitutive relation for the internal
torque $\mathbf{m}$ as
\bea
\mathbf{m} = k_1\mathbf{d}_1 + ak_2 \mathbf{d}_2 + b(k_3-k_3^{0}) \mathbf{d}_3.
\label{K3}
\eea
In the equations above,
the parameter $a$ is a measure of the bending rigidity of the cross section of
the rod. It  is the ratio $I_{1}/I_{2}$
of the two moments of inertia
of the cross-sections of the nanotube in the directions
$\mathbf{d}_1$ and $\mathbf{d}_2$.
These are conventionally oriented such that
$I_{1} \leq I_{2}$.  Hence $0 <a \leq 1$.
(For a circular cross section, $a=1$.)
$b$ is a measure of the twisting rigidity of the rod.
It is given in terms of
$a$ and the Poisson ratio $\sigma$, which
 is  a measure of the change in volume of the rod as it is
stretched:
\bea
b =2a/[(1+\sigma)(1+a)].
\label{b}
\eea
In Refs. \cite{gor1, jpha}, the term involving $k_3^{0}$
in Eq. (\ref{K3}) is absent. Hence the analysis
given there is valid for tubes
without an intrinsic twist. In the next section, we analyze
the Kirchhoff equations (\ref {K1}) and (\ref {K2})
with $k_3^{0}\neq 0$.
Our analysis will be valid for general values of $a$ and $\sigma$.

\section{Role of the
intrinsic twist}

We first analyze the static conformations.
Substituting Eqs. (\ref{g}) and (\ref {K3}) in Eqs.
(\ref{K1}) and (\ref{K2}),
we get
\bea g_{1,s} + k_2g_3 - k_3g_2 = 0,
\label{e4}
\eea
\bea g_{2,s} + k_3g_1 - k_1g_3 =0,
\label{e5}
\eea
\bea g_{3,s} + k_1g_2 - k_2g_1 = 0,
\label{e6}
\eea
\bea g_2 = k_{1,s} + (b-a)k_2k_3- b k_3^{0} k_2\,,
\label{e7}
\eea
\bea g_1 = -ak_{2,s} + (b-1)k_1k_3-b k_3^{0} k_1\,,
\label{e8}
\eea
and
\bea bk_{3,s} + (a-1)k_1k_2 = 0.
\label{e9}
\eea
As is clear from Eq. (\ref{kcomp}), the equations above
represent a
set of {\it nonlinear} coupled differential equations involving
$k$, $\tau$ and $\phi$.

To understand the role played by the intrinsic twist, we proceed as in
Refs. \cite{gor1,jpha}, and look for solutions  that correspond
to $\phi = \frac{1}{2}n \pi,\,\, n = 0,1,\ldots\,$.
Using this in Eq. (\ref{kcomp}), we see that
 Eq. (\ref{e9}) yields
\bea
\tau =\tau_0\,.
\label{tau0}
\eea
Thus the torsion of the tube, which is a measure of its
nonplanarity, is an arbitrary constant, to be determined consistently.
We focus on nontrivial conformations with
a nonvanishing $\tau_0$. Two cases arise in the analysis of
Eqs. (\ref{e4})--(\ref{e8}): \\

\noindent
Case (i):\,\,$\phi = j\pi,\,\,j = 0,1\,;\,\, k_1=0\,; \,\, k_2=(-)^{j} k$.\\
We find in this case
\bea
(b-2a)\,\tau_0 = b \,k_3^{0}\,.
\label{b2a}\eea
Using Eq. (\ref{b}) in this equation, we get
\bea
\tau_0 = - \,k_3^{0}\big/[a+\sigma(a+1)].
\label{T1}
\eea
The internal elastic force is found to be
\bea
\mathbf{g} = (-1)^{j}a
(\tau_0 \,k \,\mathbf{d}_2 -k_s\, \mathbf{d}_1)
+\left(C -\textstyle{\frac{1}{2}}a k^{2}\right)\,\mathbf{d}_3\,,
\label{G1}
\eea
where the constant $C$ represents the tension in the rod when it is
straight.\\

\noindent Case (ii):\,\,
 $\phi = (j+\frac{1}{2})\pi\,,\,\, j = 0,1\,;
\,\, k_1=(-)^j k\,;\,\, k_2=0.$  \\
We now obtain
\bea
(b-2)\,\tau_0 = b \,k_3^{0}\,.
\label{b2}
\eea
As before, using Eq. (\ref{b}) in this equation yields
\bea
\tau_0= - ak_3^{0}/[1+\sigma(a+1)].
\label{T2}
\eea
The  elastic force is determined as
\bea
\mathbf{g}= (-1)^{j}[(\tau_0 \,k \,\mathbf{d}_1+k_s \,\mathbf{d}_2)
+\left(C - \textstyle{\frac{1}{2}}k^{2}\right)\,\mathbf{d}_3\,.
\label{G2}
\eea

We now show that the intrinsic twist indeed plays a nontrivial role
in determining the torsion $\tau_0$ of the
conformation of the elastic tube.

First, if  $k_3^{0}=0$, then Eqs. (\ref{b2a}) and (\ref{b2})
yield $b=2a$ and $b=2$,
respectively, because $\tau_0 \neq 0$.
Putting in
these two values of $b$ in Eq. (\ref{b}) successively,
we get  $\sigma=-1/(1+a)$ and $\sigma=-a/(1+a)$, respectively.
Since $0 < a \leq  1$, it follows that the Poisson ratio $\sigma$
has to be {\it negative}, and in the range
$-1  <  \sigma \leq -\frac{1}{2}$ and $-\frac{1}{2} \leq \sigma < 0$,
respectively, for the two cases.
Although thermodynamic stability arguments\cite{land}
merely restrict $\sigma$ to the range
$-1 \leq \sigma \leq \frac{1}{2}$, and
 while $\sigma$ can indeed be negative for
some biopolymers\cite{gor1},
for most elastic media
one finds that  $0 < \sigma < \frac{1}{2}$.

Thus, for both cases (i) and (ii), {\it setting}
$k_3^{0}=0$ {\it determines}
the numerical value of $\sigma$, which turns
out to be negative
for any $a$. Further, $\tau_0$ can take on any arbitrary value
in this instance.

In contrast, if $k_3^{0}\neq 0$, the
corresponding torsion $\tau_0$ is
not arbitrary, but is determined in terms of $k_3^{0}$, $a$ and
$\sigma$ (which are material properties), as we
may expect on physical grounds.
This dependence can be seen from
Eqs. (\ref{T1}) and (\ref{T2}).
These equations also show that
positive as well as negative values of $\sigma$ are  allowed
in this instance,
which is a desirable feature.
We may note that when $\sigma > 0$, the torsion  $\tau_0$ and the
intrinsic twist
$k_3^{0}$ have opposite signs.
In addition,
on imposing the condition $0< a \leq 1$,
both the equations (\ref{T1}) and (\ref{T2}) lead to
the same inequality,
$\sigma < - k_3^{0}/\tau_0 \leq 2 \sigma + 1$.

\section{Exact conformations}

In the last section, we found the solutions for the torsion $\tau_0$.
Using Eq. (\ref{G1}) in Eq. (\ref{e4}), and
Eq. (\ref{G2}) in Eq. (\ref{e5}), respectively,
we can derive the nonlinear
differential equation for the curvature $k$ in
Cases (i) and (ii).  They are seen to have
the same form
\bea
k_{ss}+\textstyle{\frac{1}{2}} k^{3}=(C_2 - \tau_0^{2})\, k,
\label{koe}
\eea
where $C_2=C/a$ for Case (i), while  $C_2=C$ for Case (ii).

Note that Eq. (\ref{koe})  has
the same form as that obtained earlier
\cite{gor1, jpha}, for  the case $k_3^{0}=0$ as well,  but with the
important difference that $\tau_0$  is not arbitrary any more, but
depends on $k_3^{0}$
as shown in the last section.
Eq. (\ref{koe}) has a solution
of the form
\bea
k(s)=2(C_2-\tau_02)^{1/2}\, {\rm sech}\,\left[
(C_2-\tau_02)^{1/2}\,s\right],
\label{sech}
\eea
for $(C_{2}-\tau_{0}^{2}) > 0$.

This represents a static conformon. Now, due to  scale and
galilean invariance \cite{cole} of the
dynamic Kirchhoff equations (\ref{K1}) and (\ref{K2}),
traveling wave solution are possible for $k$, with $s$ replaced by
$(s-vt)$ in Eq. (\ref{sech}).
Here $v$ represents the velocity of the wave. This leads to a moving
conformon\cite{jpha}, which is like a solitary wave.

Now, in general, we find that the nonlinear differential equation
(\ref{koe}) supports  solutions of the form

\bea
k(s,\kappa) = 2 \sqrt{\frac{C_{2}-\tau_{0}2}{2-\kappa^{2}}}
~ {\rm dn}\, (\sqrt{\frac{C_{2}-\tau_{0}2}{2-\kappa^{2}}}~s, \kappa),
\label{dn}
\eea
where ${\rm dn}\,$ is the usual Jacobi elliptic function, with modulus
 $\kappa$ in the range $0 \le \kappa \le 1$ \cite{abra}.
For $\kappa=1$, the solution (\ref{dn}) becomes (\ref{sech}),
which was considered in Ref. \cite{jpha}.
But in contrast to that solution,
 which was spatially localized, the
solution Eq. (\ref{dn})
is a spatially  periodic function, with a finite period $2 K(\kappa)$
for all  $\kappa \ne 1$.
Here $K(\kappa)$ is the complete Jacobi integral of the first kind,
which tends to infinity for  $\kappa=1$.
Thus we call Eq. (\ref{dn}) for all $\kappa \ne 1$, a {\it conformon
lattice} solution.

In Figs. 1 to 5, we have displayed the conformations of a thin elastic
tube (rod) that correspond to
the conformon lattice solutions for the curvature,
given in Eq. (\ref{dn}), for various values of $\kappa$, and constant torsion
$\tau_0=0.5$. In all the plots,
$\sqrt {(C_2-\tau_02)}$ has been set equal to unity, since its inverse
merely scales the length.

For $\kappa=1$, the conformation is given in Fig. 1, and it corresponds
to a single loop or  conformon.
Its shape is (as expected) different from that obtained
in Ref. (\cite{jpha}) for $\tau_0=1$.
\begin{figure}[htbp]
\includegraphics{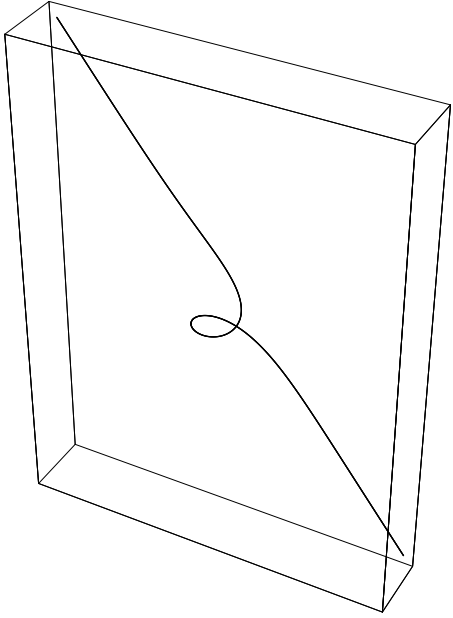}
\leavevmode
\caption
{Conformation corresponding to curvature
$k$ as given in Eq.(\ref{dn}) for $\kappa=1$.
Note the localized twisted loop formed by the axis of the tube.}
\label{fig1}
\end{figure}

We find that even a slight decrease  to  $\kappa=0.995$
leads to  a fairly large change in the conformation,
 with two loops, and a partially complete loop, as seen in Fig. 2.
\begin{figure}[htbp]
\includegraphics{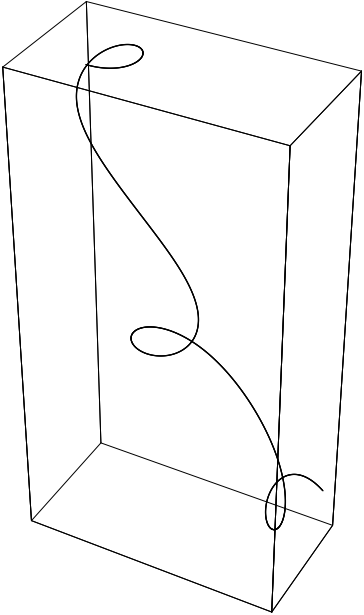}
\leavevmode
\caption
{Conformation corresponding to curvature
$k$ as given in Eq.(\ref{dn}) for $\kappa=0.995$.
Note the appearance of two loops and a partially complete loop.}
\label{fig2}
\end{figure}
From $\kappa=0.995$ to $\kappa=0.75$, there are steady changes in the
conformation.
In contrast, below  $\kappa =0.75$, the conformational
changes are much slower as $\kappa$ varies, as seen by comparing  Fig. 3
($\kappa = 0.75$) with Fig. 4  ($\kappa=0.25$), which differ very slightly.
\begin{figure}
\includegraphics{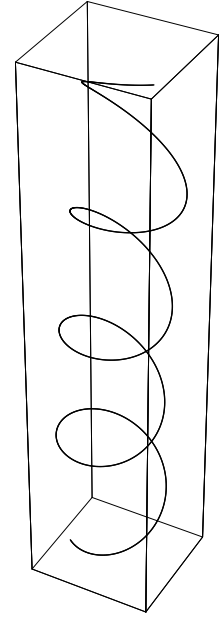}
\caption{\label{fig3}
Conformation corresponding to curvature
$k$ as given in Eq.(\ref{dn}) for $\kappa=0.75$.
There are three twisted loops of unequal sizes, and a bend.}
\end{figure}
\begin{figure}
\includegraphics{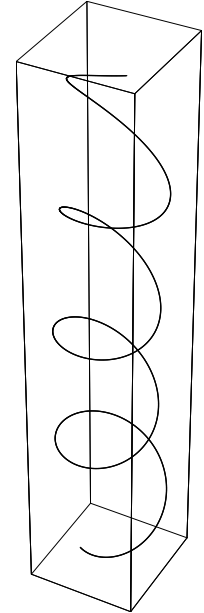}
\caption{\label{fig4}
Conformation corresponding to curvature
$k$ as given in Eq.(\ref{dn}) for $\kappa=0.25$.
There are three  twisted loops of unequal sizes, and a bend.}
\end{figure}

For  $\kappa=0$, the curvature becomes  a constant
independent of $s$. Hence this conformation
represents a  structure in which the
axis of the tube  is  coiled into a helix, with
a constant curvature and torsion. This is displayed in Fig. 5.
We parenthetically remark that while this helical tube was
essentially the only
conformation considered in
\cite{fons} in the context of the Kirchoff model, we see that
a wide class of coiled conformations corresponding to nonzero $\kappa$
get supported by the model.

\begin{figure}
\includegraphics{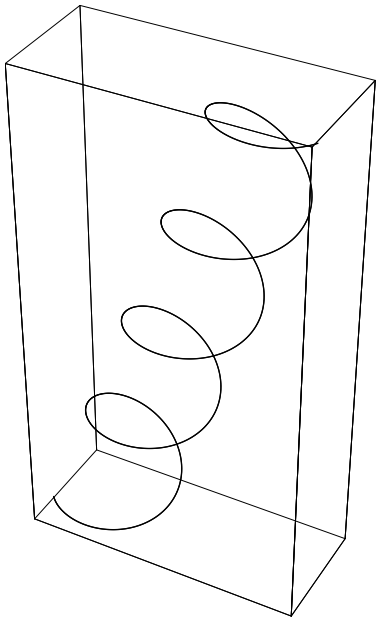}
\caption{\label{fig5}
Conformation corresponding to curvature
$k$ as given in Eq.(\ref{dn}), which becomes a constant for $\kappa= 0$.
There are four identical loops, representing  four
full windings of a helix.}
\end{figure}

As explained below the solution Eq. (\ref{sech}) for the conformon,
it is possible to have traveling wave solutions
(\ref{dn})
with $s$ replaced by $(s-vt)$.

We conclude this section with the remark that the solutions for the curvature
(Eq. (\ref{dn})) will also be applicable  for a closed tube
of length $L$ satisfying  periodic boundary conditions.
We get
$L \sqrt{\frac{C_{2}-\tau^{2}_{0}}{2-\kappa^{2}}}~
=2m K(\kappa)$, where $m$ is an integer.

\section{Quantum mechanical charge transport in a tube}

In this section, we consider the implications of the
conformon lattice solution(\ref{dn})
on  electron transport along  a nanotube. We recall some salient steps
given in \cite{jpha}, where we discussed the conformon, for the sake
of completeness.

As has been shown  \cite{gold, clar},  a thin tube
 with curvature $k(s)$ and
torsion $\tau_{0}$, induces a  quantum potential
$V_{eff}(s)=(\hbar^{2}/2\mu)[-k^{2}/4~ +~\tau_{0}2/2]$
for an electron moving on it.
Here $\mu$ is the mass of the electron.
In the time-dependent Schr\"{o}dinger equation with the
above potential, we
 can eliminate the constant torsion $\tau_{0}$
using a simple gauge transformation $\psi_{1}=\psi(s,t)\exp (-i\hbar
\tau_{0}2 t/4\mu)$.  This leads to

\bea
i\hbar \frac{\partial
}{\partial t}\psi(s,t)
=-\frac{\hbar2}{2\mu}\left(\frac{\partial2}{\partial
s2} + \frac{k2(s)}{4} \right)\psi(s,t)
\label{e13}
\eea

Using the transformation
$(t,s) \rightarrow (\frac{4\mu}{\hbar}u, \sqrt{2}s_1)$,
Eq. (\ref{e13}) becomes
\bea
i~\psi_{u}+\psi_{s_1s_1} + \frac{k2}{2}\psi =0,
\label{knls}
\eea
where $k=k(s_1)$, and the subscripts $s_1$ and $u$ stand
for the partial derivatives $\frac{\partial}{\partial s_1}$ and
$\frac{\partial}{\partial u}$.

Looking for stationary solutions of the Schr\"{o}dinger equation
(\ref{knls}) in the form
\bea
\psi(s_1,u)=k(s_1)\exp (-i E~u),
\label{e15}
\eea
 we get
\bea
\left (k_{s_{1}s_{1}} + \frac{k3}{2}\right)= -E ~k.
\label{e16}
\eea
Comparing this with Eq. (\ref{koe}), for self-consistency, we
must have
\bea
E = -(C_2 -\tau_0^{2}),
\label{E}
\eea
which is negative. We write  Eq. (\ref {e16})
in the form of a  time-independent
 Schr\"{o}dinger equation with a potential term $V$
\bea
-k_{s_{1}s_{1}} - V(s, \kappa) k = E ~k.
\label{se}
\eea
On using Eq. (\ref{dn}), we see that the potential is given by
\bea
V(s, \kappa) = -2 \frac{(C_{2} - \tau_{0}^{2})}{(2 - \kappa^{2})}
{\rm dn}\,^{2}(\sqrt{\frac{C_{2}-\tau_{0}2}{2-\kappa^{2}}}~s, \kappa),
\label{v}
\eea
which is negative as well.
Since ${\rm dn}\,^{2}(s,\kappa) \le 1$ for all $s$ and $\kappa$,
the minimum value of the potential in Eq. (\ref{e16}) is
\bea
V_{min} = -\left(C_{2}-\tau_{0}2\right )/[1 - \frac{\kappa^{2}}{2}].
\label{vmin}
\eea
Thus we have the inequality $0 \ge E \ge V_{min}$.
On using   Eq. (\ref{dn}) and Eq. (\ref{E}) in Eq. (\ref{e15}) ,  the {\it exact}
stationary solution of the time dependent
Schr\"{o}dinger equation for the electron (Eq. (\ref{knls}))
is of the form
\begin{widetext}
\bea
\psi(s_{1},\kappa,u)= 2\sqrt{\frac{C_{2}-\tau^{2}_{0}}{2-\kappa^{2}}}~
{\rm dn}\, (\sqrt {\frac{C_{2}-\tau^{2}_{0}}{2-\kappa^{2}}}~s_{1}, \kappa)~
\exp i~ (C_{2}-\tau^{2}_{0})u.
\label{psi}
\eea
\end{widetext}
Hence we see that the conformon lattice solution, a spatially periodic solution
for $k(s_{1})$ given
in Eq. (\ref{dn}),
which determines the  conformation of the elastic tube,
is also  the amplitude of the electron
wave function in (\ref{psi}).

Note that when $\kappa$ goes to $1$, the amplitude of the above
solution reduces to Eq. (\ref{sech}), which corresponds to the
static conformation in Fig. 5.
This would in turn localize the electron at the center  of the loop.

However, as we explained below Eq. (\ref{sech}),
a dynamical solution with $s$ replaced by $(s-vt)$ is  possible for the Kirchhoff equations
(\ref{K1}) and (\ref{K2}), which would describe
a moving loop.
Turning to electron transport on a tube, we  showed in \cite{jpha}
that the quantum effective potential induced by the
above (moving) curvature traps the electron, which then travels with
it along the tube.
This is
the conformon mechanism
 described in Ref. \cite{jpha} and applied to a biopolymer..

Now, we see that the conformon lattice solution for the curvature (Eq. (\ref{dn}))
to a different type of charge transport mechanism.
While for $\kappa = 1$, it was the {\it motion}  of the
conformon that led to electron transport, we see that
for $\kappa \ne 1$, {\it even} the  static conformon lattice
contributes  to  electron transport.
This happens because the electron wave function
Eq. (\ref{psi}) corresponds to a
spatially {\it delocalized state}, due to the finite periodicity
of ${\rm dn}\,$. Indeed,  Eq. (\ref{psi})
shows that since for any $\kappa$,
 ${\rm dn}\,(s,\kappa)$ does not vanish  for any $s$, the probability
of finding the electron on the polymer,
${\rm dn}\,^{2}(s,\kappa)$, is non-vanishing for all $s$.
Note that the period of ${\rm dn}\,(s,\kappa)$ increases as $\kappa$ increases,
and tends to infinity as $\kappa$ tends to $1$.  Further,
${\rm dn}\,(s,\kappa)$ itself goes to unity for all $s$ as $\kappa$
tends to zero.
The conformon lattice mechanism therefore contributes
towards enhancing electron transport along the nanotube
as $\kappa$ is decreased from $1$ to $0$.

\section{Discussion}

With regard to applications,  let us first consider
carbon nanotubes.
It has been shown using Green Function techniques in the context of
an "arm chair"  carbon nanotube that  bending of a
nanotube increases its electrical resistance\cite{roch}.
This agrees with our finding that a single bend (as in Fig. 1)
would cause localization of the electron wave function.
Thus although the
phenomenological model we have presented is a
highly idealized continuum model in the sense that it ignores
details such as electronic structure, it is able to capture the essence
of the curved geometric effects quite succinctly.
Its merits include the use of a realistic elastic model that takes into
account an intrinsic twist
and  a possible asymmetry in the bending rigidities.
We have obtained a helical conformation (Fig. 5) which has been observed
experimentally. While the fact that a helical solution {\it per se}
can be supported  by the Kirchhoff model
 is well known \cite{gor2},
here it emerges as a special case of  an elliptic function solution.
It would be interesting to experimentally
measure its  electrical resistence as a
function of the number of coils. We conjecture that the
resistance  of nanotubes where the coils appear very close to
each other may be less
than that in which the coils are far apart, since the delocalization
of the electron  will be very efficient in the former case.
Hence design of suitable experiments to verify the above would be of
interest.

More importantly, we have shown
that in addition to the helical conformation,
the Kirchhoff model  supports
shapes such as those given in Figs. 1 to 4, which have
constant torsion but  varying curvature. While we have
seen  in \cite{jpha} that a static loop conformation
(Fig. 1) localizes the charge on the loop, the new
multiloop solutions (Figs. 2 to 5) are shown to
 lead to delocalization of the wave function,
which provides  a novel mechanism for charge transport.

Conformations corresponding to closed tubes are also possible.
First, the simplest case of a circular ring  (and hence planar)
solution
is easily obtained
by setting  the curvature $k$ to be a constant $K_{0}$,
and the  torsion $\tau_{0} = 0$ in the basic equations
(\ref{e4}) to (\ref{e9}).
We  obtain $K_{0} =\sqrt{2 C_{2}}$.
This is the same as  the
 solution of $k$ found from Eq. (\ref{koe})
 by inspection in this case. In addition,
closed tube
conformations for the general elliptic function solution
(\ref{dn}) exist when its length
 $L$ is such that
$L \sqrt{\frac{C_{2}-\tau^{2}_{0}}{2-\kappa^{2}}}$
 is an integral multiple of its period , $2 K(\kappa)$,
with periodic
boundary conditions ${\bf R}(L) = {\bf R}(0)$.

In the context of DNA,  Shi and Hearst \cite{shi} have
analysed the Kirchhoff equations, by treating DNA
as a thin elastic rod with a  circular  cross-section,
 to obtain exact solutions of curvature and torsion.
They have  plotted
closed tube configurations of a
supercoiled DNA in the form of a toroidal helix.
Our analysis  and plots are for  open tubes  with
 non-circular cross-section, the helix being a special case.

Over the years, physicists have used various types of
experiments  to  measure  DNA conductivity, but the results
contradict each other.\cite{endr}
Bioscientists have employed
other types of methods\cite{boon} such as oxidative
damage in  DNA, to study charge transport.
They have  firmly established
that there is definitely an efficient transport of electrons (or holes)
through the DNA base pair stack, which proceeds over
fairly long molecular distances ($\sim 200$ \AA)
under ordinary thermal conditions.
Since the central axis  of DNA can take on curved  conformations,
 especially when it interacts with
various proteins during   replication and transcription,
 the quantum mechanical motion of
the electron  is not always  along
a straight line, but  rather, on  a curved path.
This curvature perhaps
plays a role in charge transport,
since the sensitivity of transport to
sequence-dependent conformations as well as  local
flexibilty and distortions
of the soft DNA chain have been noted  in
experiments. \cite{tran, boon}

It is indeed not easy
to measure coherent charge transport in DNA, because of effects due to
temperature, disorder, dissipation and interaction
with molecular vibrations, etc \cite{endr}. In addition,
issues connected with
the source of the charge, whether it is transport between the
leads or due to a donor-acceptor scenario, will need
careful experimental scrutiny. But we believe that
irrespective of how the charge gets injected,
given the same ambient conditions like temperature,
a coiled elastic nanotube induces a quantum potential
in its path, while
a straight tube of the same length  does not, and
this should lead to observable effects on charge transport.

We note that the
formation of DNA loops (such as in Figs. 1 to 5)
has been frequently observed \cite{matt},
and is considered
to be an important biological mechanism for regulating
gene expression.
Thus the exact conformations that we have obtained using
the elastic model indeed  seem to have  physical significance.
In view of the above, systematic experiments  should be
designed to study the
dependence of  charge transport  on  DNA  conformations.

We suggest that in
such experiments, long molecules (with length greater than the
persistence length of 50nm) which are not stretched out or attached to a
surface, but rather in their more
natural, free curved conformation (with loops, if possible)
should be used
to investigate for what kinds of conformations, there would be either
enhancement or depletion of charge migration.

In summary, as a result of an intricate interplay between classical
elasticity of the nanotube and {\it quantum} motion of the electron
in it, certain
conformations with constant torsion  are shown to lead to a class of
 corresponding exact  electronic
wave functions.  This suggests
a  charge transport mechanism that underscores the role played by
the curved geometry of the  axis of  the tube, and in particular, the
nonlinear differential equation satisfied by the curvature function.
Hence  curved geometry
should be taken into account in order
to obtain a full understanding of  charge conduction in
nanotubes.
Our work should be regarded as a first step
in this direction.

RB thanks the Council of Scientific and Industrial
Research, India, for financial support
under the Emeritus Scientist Scheme.

\end{document}